# Phonon-bound nucleon pairs predicted by the Bohm-Staver relation

J.S. Brown

*Clarendon Laboratory, 1 Parks Rd. Oxford, England*

**Abstract**

Attention is drawn to recent observation of a fluid phase of hydrogen nucleii in transition metal hydrides. It is suggested that phonon exchange processes might be responsible for the phase transition since bound pair states, analogous to Cooper pairs, would possess the required mobility and might moreover render the phase transition energetically favourable. A generic formula for the ground state binding energy based upon the results of numerical calculations is presented and a special indirect test of the theory is proposed.

**Introduction**

Neutron scattering studies of certain transition metal hydrides[1] suggest that at least some of the hydrogen nucleii exist in a delocalized state when the stoichiometric loading ratio is sufficiently high. This is surprising, since the hopping mobility of individual hydrogen nucleii in a lattice of metal ions is not nearly high enough to drive or sustain a Mott insulator-to-fluid phase transition[2]. Since superconducting electrons are considerably more mobile than conduction electrons, we were led to consider whether LA phonon exchange processes might not be responsible for the formation of hydrogen-nucleus pairs similar to Cooper pairs.

**Model outline**

The matrix element in the Fröhlich Hamiltonian for the exchange of momentum $\vec{q}$ between two particles of bare charge $Z$ and mass $m$ in a lattice of ions of mass $M$ is approximately described, in the limit of small $m/M$, by the Bohm-Staver relation:

$$\frac{Z^2 e^2}{q^2 + k_{TF}^2}\left[1 + \frac{\omega^2(q)}{[\epsilon(k+q)-\epsilon(k)]^2 - \omega^2(q)}\right]\bar{c}_{k+q} c_k \bar{c}_{k'-q} c_{k'}$$

where $\omega(q) \approx \omega_D \sin(\vec{q}\cdot\vec{a})$ (1)

This generic relation describes the combined effect of the adiabatic screening of the Coulomb repulsion due to the electron gas, characterized by Thomas-Fermi wavevector $k_{TF}$, together with the additional dielectric effect of the retarded ion vibrations characterized by the Debye frequency $\omega_D$. That the net interaction can be attractive for band state electrons close to the the Fermi surface is well-known, since this mechanism provides the foundation of the BCS theory of low temperature superconductivity. The relevance to the present case lies in the observation that the free-particle kinetic energy of an ion in a lattice $\epsilon(k)$ is typically smaller than that of band electrons by the factor $m/m_e$. The phonon-mediated retarded potential is consequently comparable to the instantaneous Coulomb interaction over a wide range of **k** and **q**. There is thus a strong *a priori* reason for suspecting the existence of paired

(light) nucleon states having binding energies considerably greater than those typical for Cooper pairs in a wide range of crystalline environments. There is also no obvious reason why such an effect should be limited to hydrogen nucleii only.

In order to calculate the properties of such pairs, it is necessary to solve a two-body Hamiltonian in which the interaction potential for the interparticle separation coordinate is given by (1). In contrast to almost any conceivable calculation involving band electrons, we can safely ignore the periodic single-particle lattice potential in 1$^{st}$ order, since we are only interested in strongly bound states. The pair is a plane-wave eigenstate of total crystal momentum $\vec{k}+\vec{k}'$ and spin $\sigma+\sigma'$ in this approximation. Whether this simplifying assumption is self-consistent will be apparent from the calculated value of the binding energy.

The two-particle Hamiltonian is still not amenable to a closed analytic solution owing to the k-dependence of the effective potential and the periodicity of the phonon dispersion curve. However, if we limit ourselves to l=0 states, the two-body Hamiltonian can be transformed into a central force problem with $r=r_1-r_2$ and spherical Bessel basis set:
$$\varphi_z(r)=\sum_{n=1}^{N} c_{zn} \sin\frac{(2\pi nr/R)}{r\sqrt{R}} \qquad (2)$$

The resultant k-space matrix can be solved by computer to an accuracy limited only by the finite value of the cut-off momentum $N/R$

With a matrix dimension N = 1000, the calculated energy values and state vectors have been found to vary smoothly over a wide range of physically realistic values for all the relevant physical parameters *viz:*

i)    Thomas-Fermi screening wavevector $k_{TF}$
ii)   crystal size $R$ expressed as a multiple of the screening distance $L_{TF}=k_{TF}R$
iii)  Debye frequency $\omega_D$
iv)   nucleon mass $m$
v)    bare nucleon charge $Z$

It has been found that all the dependencies can be expressed by the following simple generic formulae:

$$E_{GND}\approx -4\,Ze\cdot m\left[\frac{\omega_D}{k_{TF}}-\frac{4\cdot 10^{-6}}{\sqrt{L_{TF}}}\right] \quad \text{and} \quad \langle r\rangle_{GND}\approx \frac{2}{k_{TF}} \qquad (3)$$

Accordingly, given the assumptions of our model, we conclude that:

i) the ground state binding energy is in the eV range and suffices to pay for the pair ignoring the periodic lattice potential, thus validating our neglect of this in first order.

ii) the binding energy is proportional to nucleon mass $m$ so long as $m\ll M$

iii) the binding energy increases linearly with the bare charge $Ze$

iv) the binding energy increases linearly with the Debye frequency $\omega_D$

v) there exists a certain critical crystal size of the order of a few nm below which no bound state nucleon pairs can form.

vi) the radial extent of the pair is largely determined by the Thomas-Fermi screening distance and is orders of magnitude smaller than the radius of an electronic Cooper pair. The interaction between pairs can therefore be expected to be small, which provides justification for our two-body treatment. The amplitude is expected to fall exponentially at smaller separations where the Coulomb repulsion dominates.

Our calculations also revealed excited bound states, provided the crystal is large enough. The spectrum is reminiscent of that of hydrogen.

**Conclusion**

Returning to the particular case of deuterons in Pd, we obtain a crystal critical size of about $30 \, \text{Å}$ a spatial extent of about $0.4 \, \text{Å}$ and a pair binding energy of about 10 eV in the large crystal limit.

Since the bound pair wavefunction can be expected to have a much higher amplitude at the nuclear radius $r \approx 3 \cdot 10^{-15} \, m$ than any conceivable molecular or interstitial configuration, an indirect confirmation of the model described here would be the observation of $^2H + ^2H$ fusion reactions. Over nuclear timescales, the emission of high energy quanta would of course be strongly forbidden by the translational symmetry of the pair wave function, so the excited nucleons would presumably have time to dissipate their energy by emission of phonons.